\documentstyle[prl,aps,multicol]{revtex}
\renewcommand{\narrowtext}{\begin{multicols}{2} \global\columnwidth20.5pc}
\renewcommand{\widetext}{\end{multicols} \global\columnwidth42.5pc}
\input{epsf.tex}

\begin{document}
\bibliographystyle{prsty}
\draft
\title{Spectral Correlation and Response functions in Quantum Dots}
\author{N. Taniguchi$^a$~\cite{address-j}, B. D. Simons$^b$ and
  B. L. Altshuler$^{a,c}$}%
\address{%
  $^a$ Department of Physics, Massachusetts Institute of Technology, 77
  Massachusetts Avenue, Cambridge, MA 02139, USA\\ $^b$ Blackett
  Laboratory, Imperial College, Prince Consort Road, London SW 7 2BZ, UK\\
  $^c$ NEC Research Institute, 4 Independence Way, Princeton, NJ 08540, USA}
\date{\today} \maketitle

\begin{abstract}
  We derive a general relation between correlators of density of
  states fluctuations and density response functions.  It applies equally
  to quantum chaotic systems of pure symmetry (unitary, orthogonal, and
  symplectic) as well as to the crossover region between the universality
  classes.  This relation is much more robust than Wigner-Dyson
  statistics; its validity extends to disordered metals with
  finite conductance and even to the Anderson insulators with large
  localization length.
\end{abstract}

\pacs{05.45.+b,73.20.Dx,73.20.Fz}

\narrowtext
One way to investigate a many-body system is to measure its response to an
external perturbation. The response functions to which these measurements
correspond can be presented through the two-particle Green
function. In this Letter, we obtain and discuss generic
properties of the response functions of quantum chaotic systems focusing
particular attention on the behavior of disordered metallic grains
(quantum dots).  The {\em exact\/} analytical results
which we obtain apply {\em beyond\/} the universal regime described by the
random matrix theory.

Chaotic quantum dots and chaotic quantum systems in general are known to
exhibit a large degree of universality. Typically, this has been
demonstrated by focusing on the statistical properties of spectra. Perhaps
the most widely studied statistical characteristic is the two-point
correlator of density of states (DoS) fluctuations
$R_2(\omega)$~\cite{Dyson62,Mehta}.  Measuring the probability of finding
an eigenstate at a distance $\omega$ in energy from a given state, it
illustrates clearly such features as the repulsion of energy levels,
spectral rigidity, etc.
In particular, for disordered metallic systems, an expression for
$R_2(\omega)$ can be determined explicitly~\cite{Efetov83,Verbaarschot85b}
and compared with the result from the random matrix theory.
Recently the correlator was generalized to
account for spectra that disperse as a function of some external tunable
parameter
\cite{Wilkinson88,Gaspard90,Goldberg91,Simons93c,Beenakker93a}.
The parametric two-point correlator of DoS expresses the probability to
find two states separated by an amount $\omega$ and $x$ in energy and
parameter space respectively.

$R_2(\omega)$ provides just one example of a two-particle average Green
function. Others which describe responses of the system to external
perturbations involve the structure of the eigenfunctions through matrix
elements in addition to
spectral properties.
However, the striking feature to emerge from recent studies is the
observation that the different response Green functions are, in fact, not
independent, but connected by a set of differential relations
(see Ref.~\cite{Taniguchi95}, and below).
In this Letter, we will investigate the origin of these relations.  Doing
so, we will demonstrate that, for disordered quantum dots, these relations
can be substantially generalized to encompass the non-universal regime of
finite dimensionless conductance $g$. The exact analytical relations we
derive apply beyond random matrix theory, where analytical expressions for
different correlators are as yet unknown. Indeed the results are valid
even when the states are localized provided the localization length is
sufficiently large.

To make our discussion concrete, we focus on the problem of a
quantum dot subject to an external perturbation $\Phi$ whose strength is
controlled by a parameter $X$,
\begin{equation}
  H(X)={1\over 2m}\left(\bbox{p}-{e\over c} \bbox{A}\right)^2 +
  U(\bbox{r})+X\Phi (\bbox{r}).
\label{hamiltonian1}
\end{equation}
The random impurity potential $U(\bbox{r})$ is taken to be
$\delta$-correlated white-noise with $\langle U(\bbox{r}) U(\bbox{r}')
\rangle = \delta(\bbox{r}-\bbox{r}')/2\pi\nu\tau$, where $\nu$ denotes the
average DoS and $\tau$ is the relaxation time. A vector potential $\bbox{A}$
is included in the unperturbed Hamiltonian and allows an interpolation between
orthogonal (T-invariant) and unitary (broken T-invariant) symmetry.
The perturbation $\Phi(\bbox{r})$ is assumed to be local by comparison
with the mean free path $\ell$, and to preserve the generic (spin-rotation/
time-reversal) symmetries of the Hamiltonian $H(X=0)$.

\begin{figure}
%
\centerline{\epsfxsize=8cm \epsfbox{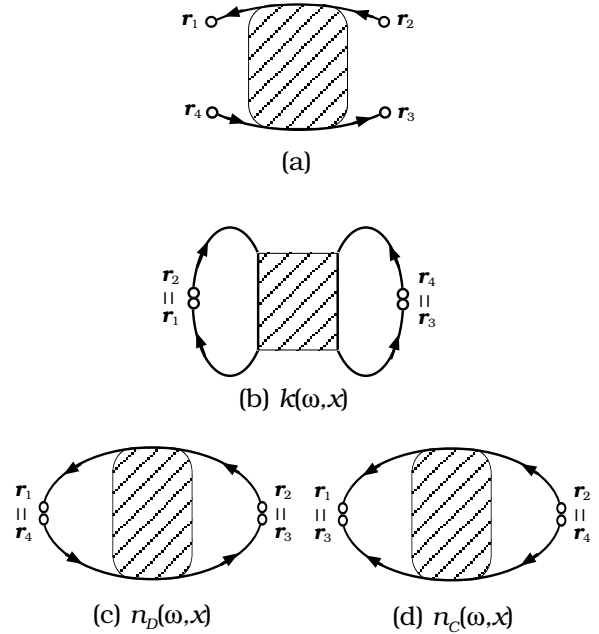}}
\caption{
  (a) Diagram representation of the two-particle Green function $\langle
  G^A(\protect\bbox{r}_1, \protect\bbox{r}_2) G^R(\protect\bbox{r}_3,
  \protect\bbox{r}_4)\rangle$.  (b) DOS
  fluctuation-type contribution $k$.  (c)~Cooperon-type contribution
  $n_C$.  (d)~Diffuson-type contribution $n_D$.
\label{fig:diagrams}}
\end{figure}

The response of the metallic grain to the external perturbation is
embodied in the two-particle Green function. The separation of fast and
slowly varying modes (see below) allow the contributions to this average
to be separated into three parts defined by three spatially
dependent functions $k$, $n_D$ and $n_C$ (Fig.~\ref{fig:diagrams}):
\widetext
\begin{eqnarray}
  &&{1\over 2\pi^2\nu^2}\left[\left\langle
  {G^A_{E_1X_1}(\bbox{r}_1,\bbox{r}_2)\,G^R_{E_2X_2}(\bbox{r}_3,\bbox{r}_4)}
\right\rangle - G_0^A(r_{12})G_0^R(r_{34})\right] \nonumber\\ &&\quad
= f(r_{12})f(r_{34})\;k(\bbox{R}_{12},\bbox{R}_{34};\Omega,X)
+ f(r_{23})f(r_{41})\;n_D(\bbox{R}_{23},\bbox{R}_{14};\Omega,X)
\nonumber \\
&&\qquad + f(r_{24})f(r_{13})\;n_C(\bbox{R}_{24},\bbox{R}_{13};\Omega,X),
\label{decomposition}
\end{eqnarray}
\narrowtext\noindent
where $G^{R,A}_{E,X}(\bbox{r}_1,\bbox{r}_2) = \langle \bbox{r}_1|
(E-H(X)\pm i0)^{-1} |\bbox{r}_2 \rangle$ denotes the Green function,
$\Omega = E_2 - E_1$, $X=X_2 - X_1$, $r_{ij}=|\bbox{r}_i- \bbox{r}_j|$,
and $\bbox{R}_{ij}=(\bbox{r}_i+\bbox{r}_j)/ 2$, and $\langle \cdots
\rangle$ represents the ensemble average.
The normalized Friedel function is defined by $f(r)
=\mbox{Im}\,G^A_0(r)/\mbox{Im}\,G^A_0(0)$ where $G^{R,A}_0(r_{12}) =
\langle G^{R,A}_{E,X}(\bbox{r}_1,\bbox{r}_2) \rangle$ denotes the average
Green function.  For a $d$-dimensional system, $f(r)$ can be presented
as a Bessel function,
\[
f(r) = \Gamma(d/2) (2/k_F r)^{d/2-1}J_{d/2-1}(k_Fr)\, e^{-r/2\ell}.
\]
We remark that our notation for the functions
$n_D$ and $n_C$ is chosen to distinguish the Diffuson from the
Cooperon-type contributions (Fig. 1). The decomposition implied by
Eq.~(\ref{decomposition}) relies on the rapid decay of $f(r)$ on a scale
comparable to the wavelength $k_F^{-1} \ll \ell$. Since the typical length
scale of $k$ and $n_{D,C}$ is the system size $L$, or the localization length
$\xi$, Eq.~(\ref{decomposition}) is applicable not only when the electronic
states are extended, but even when they are localized, provided that
$L,\xi\gg \ell$.

If the symmetry of the Hamiltonian itself is unchanged by the perturbation
$\Phi$, and $H(X)$ belongs to one of the Dyson ensembles (pure symmetry
cases), then $n_C = n_D$ for the systems in which T-invariance is
conserved (orthogonal or symplectic class), while $n_C=0$ for systems in
which T-invariance is violated (unitary class). Therefore, according to
Eq.~(\ref{decomposition}), for pure symmetries, all the statistical
properties of the two-particle Green function are embodied in the two
functions, $k$ and $n\equiv n_D$. However, remarkably they are not
independent of each other.  In particular, we will show that they are
connected by the relation
\widetext
\begin{eqnarray}
&&\int\!\! d\bbox{r}_1 d\bbox{r}_2 \left[ 2{\partial\over
    \partial X^2}+\left({\Phi(\bbox{r}_1)+\Phi(\bbox{r}_2) \over 2X}\right)
  {\partial \over \partial\Omega}\right] k(\bbox{r}_1,\bbox{r}_2;\Omega,X)
\nonumber \\
  &&\qquad = {\partial^2 \over \partial \Omega^2} \int\!\! d\bbox{r}\,
  \varphi(\bbox{r})\; \left[
  n_D(\bbox{r},\bbox{r};\Omega,X)+n_C(\bbox{r},\bbox{r};\Omega,X)\right],
\label{k-nBeyond0D}
\end{eqnarray}
\narrowtext\noindent
where
\begin{equation}
  \varphi(\bbox{r})=\int\!\! d\bbox{r}_1 d\bbox{r}_2 \Phi(\bbox{r}_1)
  f^2(r_{12}) \Phi(\bbox{r}_2) \;\delta
  (\bbox{r}-{\bbox{r}_1+\bbox{r}_2 \over 2}).
\label{def:phi}
\end{equation}
Eq.~(\ref{k-nBeyond0D}) represents the main result of the paper.
Its range of validity ($\Omega \tau, X\Phi\tau \ll 1$)
encompasses a wide interval from the quantum chaotic regime on one hand to
the localized
phase on the other.
It implies strong constraints on the properties of disordered metals
in regions where little is known about the explicit form of $k$ and $n$.

In confined or ``zero-dimensional'' cases where $g\to \infty$ and the
coordinate dependence of $k$ and $n$ can be neglected (the random matrix
limit), it makes sense to
discuss the volume averaged counterparts,
\[
\left\{\begin{array}{cc}
k(\Omega,X)\\ n_{D,C}(\Omega,X)\end{array}\right\} =
\int {d\bbox{r}_1 d\bbox{r}_2\over V^2}
\left\{\begin{array}{cc}
k(\bbox{r}_1,\bbox{r}_2;\Omega,X)\\
n_{D,C}(\bbox{r}_1,\bbox{r}_2;\Omega,X)
\end{array}\right\}.
\]
The former is related to the two-point correlator of DoS fluctuations
through the identity $R_2(\omega)=1+{\rm Re}[k(\omega,0)]$,
while the latter represents the two types of density response functions.
In the same limit an unfolding of the spectrum ($\omega =\Omega/\Delta$)
by the average level spacing $\Delta$, and a rescaling of the perturbation
$x^2=C(0) X^2$, where $C(0)=\Delta^{-2} \langle ({\partial
  E_\alpha(X)/\partial X})^2\rangle$ lifts the dependence of the
correlators on the microscopic details of the Hamiltonian including the
nature of the perturbation. This independence is a characteristic of
systems which displays quantum chaos, and the universal correlation
functions $k(\omega,x)$ and $n_{C,D}(\omega,x)$ can be applied quite
generally. In particular, studies of $n_{D,C}$ have been used to
investigate the response of chaotic systems, such as the dielectric
response in complex periodic crystals, the quantum return probability in
quantum dots~\cite{TaniguchiPrigodin}, and the statistics
of the oscillator strengths~\cite{Taniguchi95}.
In the universal limit, a direct evaluation of $k(\omega,x)$ and
$n_D(\omega,x)$ established the zero-dimensional analogue of
Eq.~(\ref{k-nBeyond0D})~\cite{Taniguchi95}.
\begin{equation}
  \label{k-n0D}
  2 {\partial\over\partial x^2}k(\omega,x) =
  {\partial^2\over\partial\omega^2} n_D(\omega,x),
\end{equation}
Note that the linear term of $\Phi$ in Eq.~(\ref{k-nBeyond0D}) is absent
since the average drift of the spectrum is excluded by $\mbox{Tr}\Phi=0$.
Since Eq.~(\ref{k-n0D}) relies on the statistical independence of
wavefunction and spectra statistics, the difference of
Eq.~(\ref{k-nBeyond0D}) and  Eq.~(\ref{k-n0D}) may imply the presence of
the correlation between the wavefunction and the spectra.

A similar result to Eq.~(\ref{k-n0D}) is found to apply beyond
zero-dimension models when $\Phi$ describes an irregular or random
potential with some finite range.  Since averaging over $\Phi$ is
expected, the linear term of $\Phi$ in Eq.~(\ref{k-nBeyond0D}) vanishes,
and $\varphi(\bbox{r})$ is independent of $\bbox{r}$.
This contrasts with the case of a Gaussian white noise random potential
when the linear term does not vanish.  In this case.
Eq.~(\ref{k-nBeyond0D}) generates a relation between the return back
probability $P(t)$ defined by
\begin{equation}
  P(t) = \int\!\!d\Omega\, e^{-i \Omega t}\langle G^A_{E,X}
  (\bbox{r}, \bbox{r}) \, G^R_{E+\Omega,X} (\bbox{r}, \bbox{r})
  \rangle .
\end{equation}
and DoS correlator $k$~\cite{Chalker95},
\begin{equation}
  2 {\partial \over \partial X^2}S(t,X=0) = -{V\overline{\varphi} \over
    2\pi^2}\; t^2 P(t),
\label{returnBackP}
\end{equation}
where we introduce the form factor $S(t,X) = \Delta^{-2}\int e^{i\Omega t}
k(\Omega,X)d\Omega $.
%
                                %

Finally, before turning to the derivation of Eq.~(\ref{k-nBeyond0D}), let
us examine what the formula implies for the relation between the universal
response functions in the crossover region between the universality classes.
If a T-invariant Hamiltonian is subject to a perturbation which violates
T-invariance, the degrees of freedom which rely on the interference of
time-reversed paths --- the Cooper channel --- are suppressed while the
Diffuson channel is unaffected~\cite{Lee85}. This is reflected in the
different behavior shown by the correlation functions $n_D$ and $n_C$.

The effect of a symmetry breaking perturbation can be demonstrated by
generalizing the Hamiltonian in Eq.~(\ref{hamiltonian1}) to include two types
of perturbation:
\begin{equation}
 H(X_o,\bbox{X}_u) = H_0 + X_o \Phi_o + \bbox{X}_u\cdot\bbox{\Phi}_u,
\end{equation}
where $\Phi_o$ preserves T-invariance, whereas $\bbox{\Phi}_u =
\{\Phi^{(i)}_u\}$ breaks
it.  Since there are, in principle, many different ways to violate
T-invariance, we designate $\bbox{\Phi}_u$ as a vector
describing different possible perturbations and assume
$\mbox{Tr}\ [\Phi_u^{(i)} \Phi_u^{(j)}] \propto \delta_{ij}$.
In general, the two-particle average Green function $\langle
G^R_{E_1,X_{o1}, \bbox{X}_{u1}}\, G^A_{E_2,X_{o2},\bbox{X}_{u2}}\rangle$
depends, not on each of the many parameters, but only on five: the energy
difference, $X_{\beta}= X_{\beta 2}-X_{\beta 1}$ ($\beta =u,o$),
$|\bbox{\bar X}|=|\bbox{X}_{u2}+ \bbox{X}_{u1}|/2$ and an angle $\theta$
specifying the ``relative direction,'' $\bbox{\bar X}\cdot\bbox{X} =
|\bbox{\bar X}|| \bbox{X}| \cos\theta$~\cite{Taniguchi94}.

In the universal or random matrix limit, the rescalings
\begin{mathletters}
\begin{eqnarray*}
  &&x_o^2 = C_o(0;0) X_o^2,\: x_u^2 = C_u(0;\infty )
  X_u^2,\:\bar x^2 = C_u(0;\infty ) \bar X^2\\
  &&C_{\beta}(0;\bar X) ={1\over\Delta^2} \left\langle \left({\partial
    E_\alpha/\partial X_\beta}\right)^2 \right\rangle,\quad (\beta=o,u),
\end{eqnarray*}
\end{mathletters}\noindent
lead to a generalization of Eq.~(\ref{k-n0D}) for the crossover interval
between orthogonal and unitary ensembles:
\widetext
\begin{equation}
  \left\{ {2{\partial \over {\partial (x_o^2)}}\pm \left[ {{\partial \over
          {\partial (x_u^2)}}-{1 \over 4}{\partial \over {\partial (\bar
            x^2)}}-{\cos\theta\over 2}\left( {{1 \over {4\bar x^2}}-{1
            \over {x_u^2}}} \right){\partial \over {\partial\cos\theta }}}
    \right]} \right\}\;k={1\over 2}{{\partial ^2} \over {\partial \omega
      ^2}}n_{D,C}.
\label{k-nCrossover}
\end{equation}
\narrowtext\noindent
Since $k(\omega,x_o,\bar x,x_u,\cos \theta)$ is known
explicitly~\cite{Taniguchi94}, this formula determines the response
functions $n_C$, $n_D$ within the crossover region.

We now outline the main elements involved in the derivation together with
that of the decomposition formula of Eq.~(\ref{decomposition}). For
brevity we will confine our discussion to the properties of T-invariant
systems in which the symmetry is not violated by the perturbation. An
extension to the crossover regime can be achieved straightforwardly. Our
approach and discussion is based on the field theoretic construction
developed to describe the low energy ``excitations'' or diffusion modes in
disordered metals, where statistical properties of two-point correlators
are presented through an effective nonlinear $\sigma$-model involving
$8\times 8$ supermatrix fields $Q$~\cite{Efetov83}.

Following the conventions of Ref.~\cite{Efetov83}, the $8\times 8$
supermatrix can be decomposed into $4\times 4$ submatrices $Q^{ab}$
($a,b=1,2$) which reflect the advanced and retarded subspace.
The spatially dependent correlations functions can then be expressed as
\begin{mathletters}
\label{def:knDnC}
\begin{eqnarray}
  &&2 k(\bbox{r}_1,\bbox{r}_2;\Omega,X) = -1-\left\langle
  Q^{11}_{33}(\bbox{r}_1) Q^{22}_{33}(\bbox{r}_2) \right\rangle _Q,
\label{def:k}\\ %
  &&2n_D(\bbox{r}_1,\bbox{r}_2;\Omega,X) = -\left\langle
  Q^{12}_{33}(\bbox{r}_1) Q^{21}_{33}(\bbox{r}_2) \right\rangle_Q,
\label{def:nD}\\%
  &&2n_C(\bbox{r}_1,\bbox{r}_2;\Omega,X) = -\left\langle
  Q^{12}_{34}(\bbox{r}_1) Q^{21}_{43}(\bbox{r}_2) \right\rangle_Q.
\label{def:nC}
\end{eqnarray}
\end{mathletters}
\noindent where the integration is restricted to the manifold on which
$Q^2=\openone_8$ and $\left\langle \cdots \right\rangle _Q =\int
{DQ\,\left( \cdots\right)}\,e^{-F[Q]}$.
The effective action $F[Q]$ is given by
\begin{eqnarray}
  && F[Q]={\pi\nu\over 8}\int\!\! d\bbox{r}\, \mbox{STr}\left\{ D (\nabla
  Q)^2 + 2i[\Omega -X\Phi(\bbox{r})] {Q\Lambda }\right\}
  \nonumber\\ %
  &&\qquad -{\pi^2 \nu^2 \over 16}X^2\int\!\! d\bbox{r}
   \varphi(\bbox{r})\mbox{STr}[Q\Lambda Q\Lambda],
\label{FreeEnergy}
\end{eqnarray}
where $D=v_F^2 \tau/d$ denotes the diffusion coefficient, and $\Lambda
=\mbox{diag}(\openone_4,-\openone_4)$ is the constant matrix that breaks
the  symmetry  between the retarded and advanced Green functions.
The explicit structure of $Q$ corresponding to the
three universality classes can be found in Ref.~\cite{Efetov83}.

The decomposition formula of Eq.~(\ref{decomposition}) is obtained by
presenting the average two-particle Green function in terms of the
supermatrix Green functions,
\begin{eqnarray}
  &&\left\langle
  {G^A_{E_1,X_1}(\bbox{r}_1,\bbox{r}_2)\,G^R_{E_2,X_2}(\bbox{r}_3,\bbox{r}_4)}
\right\rangle = \left\langle {\cal
  G}^{11}_{33}(\bbox{r}_1,\bbox{r}_2)\,{\cal
  G}^{22}_{33}(\bbox{r}_3,\bbox{r}_4)\right\rangle_Q\nonumber\\
&&\ + \left\langle {\cal G}^{12}_{33}(\bbox{r}_1,\bbox{r}_4)\,{\cal
  G}^{21}_{33}(\bbox{r}_3,\bbox{r}_2) + {\cal
G}^{12}_{34}(\bbox{r}_1,\bbox{r}_3)\,{\cal
G}^{21}_{43}(\bbox{r}_4,\bbox{r}_2)\right\rangle_Q,\nonumber\\
\label{g-calG}
\end{eqnarray}
where ${\cal G}=[E-\bbox{p}^2/2m -(\Omega-X\Phi) \Lambda/2 -
iQ/2\tau]^{-1}$.
For small values of the parameters, the explicit dependence of ${\cal G}$ on
$\Omega$ and $X$ can be neglected. Taking account of the slow variation of
$Q(\bbox{r})$, ${\cal G}$ can be separated into fast and slowly varying
spatial modes:
\begin{equation}
  {\cal G}(\bbox{r}_1,\bbox{r}_2) \approx G^A_0(r_{12}) + i\pi\nu
  f(r_{12}) [Q(\bbox{R}_{12})-\Lambda].
\label{calG-Q}
\end{equation}
Substitution of Eqs.~(\ref{g-calG},\ref{calG-Q}) into Eq.~(\ref{def:knDnC})
leads to Eq.~(\ref{decomposition}).

To obtain Eq.~(\ref{k-nBeyond0D}), we examine the
derivative
\begin{equation}
{\partial \over \partial X} k(\Omega,X)
= \int \!\!d\bbox{r} d\bbox{r}_1 d\bbox{r}_2\, (\delta k_1
  + \delta k_2),
\end{equation}
where
\begin{eqnarray*}
  &&\delta k_1 = {i\pi\nu\over 8} \left\langle \Phi(\bbox{r})
  \mbox{STr}[Q(\bbox{r})\Lambda] Q^{11}_{33}(\bbox{r}_1)
  Q^{22}_{33}(\bbox{r}_2) \right\rangle_Q,\\ &&\delta k_2 =
  -{\pi^2\nu^2 X\over 16}\left\langle \varphi(\bbox{r})
  \mbox{STr}[(Q(\bbox{r})\Lambda)^2] Q^{11}_{33}(\bbox{r}_1)
  Q^{22}_{33}(\bbox{r}_2) \right\rangle_Q.
\end{eqnarray*}
Since $\delta k_1$ is given through $\Phi(\bbox{r}) (\delta/\delta
\Phi(\bbox{r})) k(\bbox{r}_1,\bbox{r}_2;\Omega,X)$, the first term
can be recast in the form
\begin{eqnarray}
&&{-1\over 2}\int
  d\bbox{r}_1 d\bbox{r}_2 \left( \Phi(\bbox{r}_1) {\partial \over \partial
    E_1}-\Phi(\bbox{r}_2) {\partial \over\partial E_2}\right)
  k(\bbox{r}_1,\bbox{r}_2) \nonumber\\ %
&&=-\int
  d\bbox{r}_1 d\bbox{r}_2 \left( {\Phi(\bbox{r}_1)+\Phi(\bbox{r}_2)\over
    2}\right){\partial \over \partial\Omega}
  k(\bbox{r}_1,\bbox{r}_2). \label{deltak1} %
\end{eqnarray}
To calculate $\delta k_2$, we work with the symmetrized expression for $k$
and $n_{D,C}$ in the fermion-boson subspace~\cite{Zirnbauer86b},
\begin{eqnarray}
  && k= -{1 \over 2}-{1\over 32}
  \left\langle\mbox{STr}[k_s Q^{11}(\bbox{r}_1)] \mbox{STr}[k_s
    Q^{22}(\bbox{r}_2)]\right\rangle,\\
    &&n_D+n_C = {1\over 16}\left\langle\mbox{STr}[k_s
    Q^{12}(\bbox{r}_1)k_s Q^{21}(\bbox{r}_2)]\right\rangle_Q,
\end{eqnarray}
where $k_s=\mbox{diag}(\openone_2,-\openone_2)$,
and utilize the identity,
\begin{eqnarray}
  &&\left\langle
  \mbox{STr}[Q^{12}Q^{21}(\bbox{r})]\mbox{STr}[k_s
  Q^{11}(\bbox{r}_1)]\mbox{STr}[k_s Q^{22}(\bbox{r}_2)]\right\rangle_Q
  \nonumber\\ && =\left\langle\mbox{STr}[Q^{11}(\bbox{r}_1)]
  \mbox{STr}[Q^{22}(\bbox{r}_2)] \mbox{STr}[k_s Q^{12} k_s
  Q^{21}(\bbox{r})]\right\rangle_Q.
\label{Qidentity}
\end{eqnarray}
The identity originates from the freedom to choose the
fermion or boson field to calculate correlators, i.e., the fermion-boson
supersymmetry of the $\sigma$-model. (This property was presented
explicitly for the two-point correlator of $Q$-matrices in
Ref.~\cite{Zirnbauer86b}.) Note that since this fermion-boson symmetry is
required to impose the correct normalization, it is present even in the
crossover interval between universality classes.
Since $\mbox{STr}[Q^{11}]=-\mbox{STr}[Q^{22}]$ by the saddle
point condition on the $Q$-matrix, the right hand side of
Eq.~(\ref{Qidentity}) coincides with the second derivative of $n_D+n_C$
with respect to $\Omega$.
In this way, the second integral over $\delta k_2$ can be shown to be
equal to the right hand side of Eq.~(\ref{k-nBeyond0D}).

We emphasize that, since the proof of Eq.~(\ref{k-nBeyond0D}) relies only
on the supermatrix nonlinear $\sigma$-model, it applies with the same
generality.  This implies that the identity encompasses the behaviors both
of the mobility edge as well as in the localized phase.

In conclusion, we have derived and discussed a relation between spectral
correlators and response functions in disordered metals which applies over
a range which encompasses quantum chaos (random matrix limit), the
metallic phase (large conductance) and even the Anderson insulating regime
with a large localization length.


We are grateful to A. V. Andreev, and I. V. Lerner for useful discussions.
The work of NT and BDS was supported in part by the Joint Services
Electronic Program No. DAAL 03-89-0001 and by NSF grant No. DMR 92-14480.
BDS is grateful for the support of the Royal Society through a Research
Fellowship.

%
%


\begin{references}

\bibitem[*]{address-j}
On leave from Department of Applied Physics, Tokyo University, Tokyo, Japan.

\bibitem{Dyson62}
F.~J. Dyson, J. Math. Phys. {\bf 3},  140  (1962).

\bibitem{Mehta}
M.~L. Mehta, {\em `Random Matrices --- Revised and Enlarged Second Edition'}
  (Academic Press Inc., San Diego, 1991).

\bibitem{Efetov83}
K.~B. Efetov, Adv. Phys. {\bf 32},  53  (1983).

\bibitem{Verbaarschot85b}
J.~J.~M. Verbaarschot, H.~A. Weidenm\"{u}ller, and M.~R. Zirnbauer, Phys. Rep.
  {\bf 129},  367  (1985).

\bibitem{Wilkinson88}
M. Wilkinson, J. Phys. A {\bf 21},  4021  (1988).

\bibitem{Gaspard90}
P. Gaspard, S.~A. Rice, M.~J. Mikeska, and K. Nakamura, Phys. Rev. A {\bf 42},
  4015  (1990).

\bibitem{Goldberg91}
J. Goldberg {\it et~al.}, Nonlinearity {\bf 4},  1  (1991).

\bibitem{Simons93c}
B.~D. Simons and B.~L. Altshuler, Phys. Rev. B {\bf 48},  5422  (1993).

\bibitem{Beenakker93a}
C.~W.~J. Beenakker, Phys. Rev. Lett. {\bf 70},  4126  (1993).

\bibitem{Taniguchi95}
N. Taniguchi, A.~V. Andreev, and B.~L. Altshuler, Europhys. Lett. {\bf 29},
  515  (1995).

\bibitem{TaniguchiPrigodin} N. Taniguchi and B.~L. Altshuler, Phys. Rev.
  Lett. {\bf 71}, 4031 (1993); V.~N. Prigodin, B.~L. Altshuler, K.~B.
  Efetov, and S. Iida, Phys. Rev. Lett.  {\bf 72}, 546 (1994).

\bibitem{Chalker95}
J. T. Chalker, I. V. Lerner and R. Smith, (unpublished).

\bibitem{Lee85}
P.~A. Lee and T.~V. Ramakrishnan, Rev. Mod. Phys. {\bf 57},  287  (1985).

\bibitem{Taniguchi94} Note that $\theta$ is assumed to be $0$ in N.
  Taniguchi, A. Hashimoto, B.~D. Simons, and B.~L.  Altshuler, Europhys.
  Lett.  {\bf 27}, 335 (1994).

\bibitem{Zirnbauer86b}
M.~R. Zirnbauer, Nucl. Phys. {\bf B265[FS15]},  375  (1986).

\end{references}


%

\widetext

\end{document}